\documentclass{emulateapj}
\usepackage{apjfonts}
\usepackage{amsmath}

\newcommand{\myemail}{kmurase@yukawa.kyoto-u.ac.jp}

\newcounter{ichi}
\newcounter{ni}
\newcounter{san}
\slugcomment{Submitted}

\shorttitle{Implications of UHECRs for Transient Sources}
\shortauthors{Murase \& Takami}

\begin{document}


\title{Implications of Ultra-High-Energy Cosmic Rays for Transient Sources in the Auger Era}


\author{Kohta Murase$^{1}$ and Hajime Takami$^{2}$}


\altaffiltext{1}{YITP, Kyoto University, Kyoto, Oiwake-cho,
Kitashirakawa, Sakyo-ku, Kyoto, 606-8502, Japan; \myemail}
\altaffiltext{2}{Department Physics, School of Science, the University
of Tokyo, 7-3-1 Hongo, Bunkyo-ku, Tokyo 113-0033, Japan}


\begin{abstract}
We study about ultra-high-energy cosmic rays (UHECRs) from
transient sources, propagating in the Galactic and intergalactic
space.
Based on the recent observational results, we also estimate upper and 
lower bounds on the rate of transient UHECR sources and required isotropic 
cosmic-ray energy input per burst as
$0.1~{\rm Gpc}^{-3}~{\rm yr}^{-1} \lesssim \rho_0 \lesssim 
{10}^{3.5}~{\rm Gpc}^{-3}~{\rm yr}^{-1}$ and 
${10}^{49.5}~{\rm ergs} \lesssim 
\tilde{\mathcal{E}}_{\rm HECR}^{\rm iso} \lesssim 
{10}^{54}~{\rm ergs}$, through constraining the 
apparent burst duration, i.e., dispersion in arrival times of UHECRs.
Based on these bounds, we discuss implications for proposed 
candidates such as gamma-ray bursts and active galactic nuclei.
\end{abstract}


\section{\label{sec:level1}Introduction}
The origin of ultra-high-energy cosmic rays (UHECRs) is one of the biggest 
mysteries in astroparticle physics. So far, a number of possibilities 
were proposed, and several acceleration mechanisms have been 
theoretically developed \citep[see, e.g.,][and references there in]{Kac08}. 
However, physical conditions in these 
potential sources are uncertain, and observational progress for source 
identification has been limited by the scarcity of experimental data at the 
highest energies \citep[see, e.g.,][]{NW00}. 

The recent observational results of large area detectors such as 
the Akeno Giant Air Shower Array (AGASA), High Resolution Fly's Eye (HiRes), 
and especially the Pierre Auger Southern Observatory 
(PAO), have started to give us crucial clues to the association of 
UHECRs with UHECR sources.  
Indeed, the first results of the PAO reported a significant correlation 
between the arrival directions of the highest-energy cosmic rays with the 12th 
Veron-Cetty \& Veron catalog of active galactic nuclei (AGNs) closer 
than 75 Mpc \cite{Abrea07,Abrea08}. 
Although this result has not been confirmed by the HiRes \cite{Abbea08}  
and criticized by several authors \cite{Gorea07}, 
it has also received some confirmations, and it would be an important step 
towards solving the UHECR mystery \citep[see, e.g.,][]{Sta08}. 

However, one should not overinterpret the significance of these 
results. Although several authors also reported correlations of UHECRs 
with AGNs \cite{Geoea08,Mosea08,Zawea08}, 
one cannot exclude the possibility of other objects associated with
the large scale structure of the universe, 
which is inhomogeneous up to dozens of Mpc. 
Significant correlations of UHECRs with galaxies can also be found 
\cite{KW08,Ghiea08,Takea08}, so that gamma-ray bursts (GRBs) 
\cite{Vie95,Wax95a,MINN06} and magnetars \cite{Aro03} can be sources.  
 
Even if the association of 
UHECRs with AGNs is real, the 
report by the PAO brought us several new questions on the nature of AGNs 
generating UHECRs. Surprisingly, the large majority of the correlating 
AGNs seems radio-quiet, a class of objects not showing 
any nonthermal high-energy emission in their photon spectrum \cite{Geoea08}. 
Radio-loud AGNs, showing high-energy nonthermal emission, are more plausible 
candidates in the conventional jet paradigm \citep[e.g.,][]{RB93,NMA95}. 
Although the association with them is argued 
\cite{Mosea08}, it seems that the power of the correlating 
AGNs are insufficient to produce UHECRs \cite{Zawea08}. 
The above problem may be solved if UHECRs are produced during 
intense but short-duration flares 
\cite{FG08}. 
The magnetic fields in the universe deflect UHECRs, so that 
UHECRs are significantly delayed compared to photons and  
neutrinos generated during the bursts \cite{MW96}.  
A transient hypothesis might also help to reproduce the 
isotropy of the arrival distribution of UHECRs 
at $\sim 10^{19}$ eV \cite{TS08b}. 

In this letter, we focus on the possibility that UHECR sources are 
transient, and evaluate the deflection angles and arrival times of UHECRs
through numerical calculations, considering both of the Galactic magnetic field (GMF) and 
intergalactic magnetic field (IGMF). The required cosmic-ray energy input 
and rate of the sources are estimated. 
In this work, UHECRs are also assumed to consist of protons. 

\section{Propagation and Characteristics of UHECRs from Transient Sources}
We briefly describe the method of calculation and 
characteristics of UHECRs from transient sources. 
UHECRs ejected from their sources are deflected by the GMF and IGMF 
during their propagation. 
Only if the deflection angle $\theta_d (E,D)$ is small, 
where $E$ is the energy of UHECRs at the Earth and $D$ is the source distance, 
we could see a positional correlation of the highest-energy events 
with the sources at observationally suggested small-angle separations.   
The deflection also causes the time delay $t_d (E,D)$ between arriving 
times of an UHECR and a light emitted at the same time. 
UHECRs with the same energy have different arrival times, 
because of not only different particle trajectories 
but also stochastic photomeson production \cite{MW96}. 
Therefore, the time delay has certain distribution 
with an averaged delayed time $\bar{t}_{d} (E,D)$ and 
standard deviation in arrival times $\sigma_{d}(E,D)$,  
The arrival time spread 
$\sigma_{d}$ can be regarded as the apparent duration of an UHECR burst. 

Clearly, the magnetic fields play an essential role on both of 
$\theta_d$ and $\sigma_d$. 
For intergalactic propagation, one can typically expect 
$\sigma_{d} \sim {\bar{t}}_{d} \approx \frac{D \theta_d^2}{4c} \simeq 
{10}^{5}~{\rm{yrs}}~ E_{20}^{-2} D_{100 \, \rm{Mpc}}^{2} 
B_{\rm{IG},-9}^2 \lambda_{\rm{Mpc}}$ \cite{MW96}, 
which is also confirmed by our numerical calculations. 
Due to limited statistics of the highest-energy events, it is 
convenient to use quantities weighted by the observed cosmic-ray spectrum. 
The apparent burst duration of UHECRs above the threshold energy 
$E_{\rm{th}}$ is   
\begin{equation}
\tau_d (>E_{\rm{th}}) = \frac{1}{\mathcal{N}_0} 
\int_{E_{\rm th}}^{\infty} dE \frac{d \mathcal{N}_0}{dE}(E) 
\frac{\int_0^{D_{\rm max}(E)} dD D^2  \sigma_{\rm d} (E,D)} 
{\int_0^{D_{\rm max}(E)} dD D^2}, 
\end{equation}
where $d \mathcal{N}_0/dE$ is the UHECR spectrum observed at the Earth, 
$\mathcal{N}_0 = \int_{E_{\rm th}}^{\infty} dE 
\frac{d \mathcal{N}_0}{dE}(E) $ is the 
normalization factor, and $D_{\rm max}(E)$ is the maximum distance 
of UHECRs that can reach the Earth at the energy $E$. 
In this work, we adopt $E_{\rm th}= {10}^{19.75}$ eV as the 
threshold energy, according to the PAO results. 

Through $\tau_d$, we can relate the local rate of transient sources 
$\rho_0$ with the apparent source density $n_s$. We have 
\begin{eqnarray}
n_s (>E_{\rm th}) = \frac{1}{\mathcal{N}_0} \int_{E_{\rm th}}^{\infty} 
dE \frac{d \mathcal{N}_0}{dE}(E) n_0(E), 
\end{eqnarray}  
where 
\begin{equation}
n_0(E) \approx \rho_0 \frac{\int_0^{D_{\rm max}(E)} dD D^2 
\sigma_d (E,D)}{\int_0^{D_{\rm max}(E)} dD D^2}. 
\end{equation} 
Note that $n_s$ can be estimated from the observed small scale anisotropy 
in the arrival distribution of the highest-energy cosmic rays with 
energies above $E_{\rm th}$.  
For example, the small scale anisotropy observed by the AGASA implied 
$n_s \sim 10^{-6}$-$10^{-4}~{\rm Mpc}^{-3}$ \citep[e.g.,][] 
{Yosea03,KS05,TS07}. 
The more recent PAO data imply $n_s \sim 10^{-4}~{\rm Mpc}^{-3}$ 
\cite{TS08b}, and we hereafter adopt this value. 
Then, the local burst rate is estimated via $\rho_0 \approx n_s/\tau_d$. 

Assuming that the sources are uniform, 
we can also estimate typical values of the isotropic cosmic-ray energy 
input per burst at the energy $E$ 
as $\tilde{\mathcal{E}}_{\rm{CR}}^{\rm{iso}} (E) 
\approx  E^2 \frac{d \dot{N}_{\rm{CR}}}{d E} (E)/\rho_0$. 
Here, $E^2 \frac{d \dot{N}_{\rm{CR}}}{d E} (E)$ is the UHECR energy budget 
per volume per year at the energy $E$. Through our numerical 
calculations, we obtain $E^2 \frac{d \dot{N}_{\rm{CR}}}{d E} 
({10}^{19}~{\rm eV}) \simeq (0.5-2) \times 
{10}^{44}~{\rm ergs}~{\rm Mpc}^{-3}~{\rm yr}^{-1}$ (depending on 
the source spectral index $s$) from the PAO data 
\citep[see also, e.g.,][]{Wax95b,BGG06}. 

The thing left to do is to calculate the distribution of deflection 
angles of arrival times. Our method of calculation, taking into 
account the GMF as well as the IGMF, is described as below. 
The IGMF strength $B_{\rm{IG}}$ is very uncertain, but 
we can estimate upper bounds on $B_{\rm IG}$ and resulting $\tau_d$, 
by comparing the calculated distribution of deflection angles to the 
typical angular separation of observed UHECRs. In this work, we adopt 
$\psi \sim 5^{\circ}$ as the angular separation, according to the PAO 
results \cite{Abrea08,Takea08}. 
As the energy distribution of cosmic rays at a source, 
we assume power-law spectra of $dN/dE_g \propto E_g^{-s}$.  

\subsection{Propagation in the Galactic Space}
Propagation in the Galactic space is important for the deflection of UHECRs. 
The corresponding delay time would typically be smaller than that by the IGMF, 
but it is not zero 
and is unavoidable. 
Under a given separation angle, 
the coherent component of the GMF leads to the minimum delay time, 
and the lower bound on $\tau_d$ is also obtained. 
Following the method used in Takami \& Sato 2008a, we pursue 
cosmic-ray trajectories with proton mass and charge of -1 from the 
Earth, and calculate their delay times for given GMF models. 
We define a sphere with the radius of 40 kpc, centered of the Galactic 
center, as the boundary of Galactic space. 
As a GMF model \citep[see, reviews,][]{Val04,Han07}, 
a bisymmetric spiral field with the even 
parity is adopted. This leads to conservative estimate of $\tau_d$, 
although other models such as axisymmetric spiral field  
may be possible due to uncertainty in the GMF \cite{Val05}. 
Note that all the energy loss processes can be neglected for 
propagation in the Galactic space. 

The delay time and arrival time spread by the GMF depend on arrival 
directions of UHECRs, where the averaged standard deviation of 
observed $k$ events is $\sigma_d = (1/k) \Sigma_i \sigma_{d,i}$. In 
this work, we instead use $\sigma_{d} (E) = \frac{1}{4\pi} \int d\Omega \, 
\frac{d \sigma_{d}}{d \Omega} (E,\Omega)$. 
Through our numerical calculations, we found that this gives us 
reasonable estimate of $\sigma_d$, even though 
the GMF leads to the hole in arrival directions of UHECRs \cite{TS08a}.     
In Table 1, we show resulting lower bounds on $\tau_d$. Here
$p~{(<0)}$ is the pitch angle of the spiral component of the GMF 
at the vicinity of the solar system, 
and smaller values of $-p$ leads to smaller deflection angles of UHECRs.

\begin{table}[tb]
\begin{center}
\caption{\label{Tab1} 
Upper bounds on $\tau_d^{-1}$ from the GMF without a dipole field.}
\begin{tabular}{|c||c|c|}
\hline $s$ & $\tau_d^{-1}$ [$\rm{yr}^{-1}$] for $p=-10^{\circ}$ 
& $\tau_d^{-1}$  [$\rm{yr}^{-1}$] for $p=-8^{\circ}$\\
\hline
\hline 2.0 & $2.5 \times {10}^{-2}$ & $3.1 \times {10}^{-2}$\\
\hline 2.2 & $2.5 \times {10}^{-2}$ & $3.0 \times {10}^{-2}$\\
\hline 2.4 & $2.5 \times {10}^{-2}$ & $3.0 \times {10}^{-2}$\\
\hline 2.6 & $2.4 \times {10}^{-2}$ & $2.9 \times {10}^{-2}$\\
\hline
\end{tabular}
\end{center}
\end{table}

The vertical magnetic field near the solar system 
and many 
gaseous filaments perpendicular to the Galactic plane 
are observed, indicating another regular component \cite{Han07}. 
Also, a dipole field with the odd party is predicted by the dynamo theory. 
Hence, we also consider cases of the GMF with a 
dipole magnetic field whose strength is normalized to 0.3 $\mu$G at 
the vicinity of the solar system.  
In Table 2, resulting lower bounds on $\tau_d$ are shown. 
However, note that there is no direct observational evidence of such a
dipole field. 
\begin{table}[tb]
\begin{center}
\caption{\label{Tab1} 
Upper bounds on $\tau_d^{-1}$ from the GMF with a dipole field.}
\begin{tabular}{|c||c|c|}
\hline $s$ & $\tau_d^{-1}$ [$\rm{yr}^{-1}$] for $p=-10^{\circ}$ 
& $\tau_d^{-1}$  [$\rm{yr}^{-1}$] for $p=-8^{\circ}$\\
\hline
\hline 2.0 & $6.4 \times {10}^{-4}$ & $6.1 \times {10}^{-4}$\\
\hline 2.2 & $6.3 \times {10}^{-4}$ & $6.0 \times {10}^{-4}$\\
\hline 2.4 & $6.2 \times {10}^{-4}$ & $5.9 \times {10}^{-4}$\\
\hline 2.6 & $6.2 \times {10}^{-4}$ & $5.9 \times {10}^{-4}$\\
\hline
\end{tabular}
\end{center}
\end{table}

\subsection{Propagation in the Extragalactic Space}
Propagation in the extragalactic space is expected 
to play an essential role on both of the 
deflection and time delay of UHECRs. 
We numerically calculate the distribution of $\theta_d$ and $t_d$ 
for given IGMFs. Our method of calculation is similar to that used 
in Yoshiguchi et al. 2003, where proton propagation is treated 
as Monte Carlo simulations. We set 10 logarithmic bins per 
logarithmic energy interval, and isotropically inject 5000 protons 
for every energy bin. Particle trajectories are pursued at every 1 
Mpc for $D<100$ Mpc while at every 10 Mpc for $D>100$ Mpc. 
As relevant energy loss processes, 
we consider photomeson production and Bethe-Heitler processes with the 
cosmic microwave background photons, 
and the adiabatic energy loss due to the expanding universe.  
The resulting distribution of $\theta_d$ or $t_d$ is 
\begin{equation}
f_d (E,D) = \int_{E_g^{\rm{min}}}^{\infty}dE_g~{\tilde{f}}_d(E,D;E_g), 
\end{equation}
where $E_g^{\rm{min}}(E_{\rm th},D)$ is the minimum energy at a source 
with the distance $D$, of protons observed with $E_{\rm th}$ at the 
Earth. $\tilde{f}_d(E,D;E_g)$ is the more basic distribution of 
$\theta_d$ or $t_d$, generated by cosmic rays with $E_g$ at a source. 

In this work, we consider an uniform turbulent IGMF with the  
Kolomogorov turbulence spectrum as the extragalactic magnetic field.
Although the IGMF is highly uncertain, we can constrain it by 
comparing the calculated $\theta_d$ distribution to $\psi$. 
As a result, we found $B_{\rm{IG}} \lambda_{\rm{coh}}^{1/2} \lesssim 
\rm{nG} \, \rm{Mpc}^{1/2}$ is required for the averaged deflection 
angle $\bar{\theta}_d$ not to exceed the typical angular 
separation $\psi \sim 5^{\circ}$. 
It is consistent with the result from Faraday rotation measurements 
\cite{Kro94}, but independently obtained. 
Hence, 
$\tau_d$ obtained for 
$B_{\rm{IG}} \lambda_{\rm{coh}}^{1/2} \sim 
\rm{nG} \, \rm{Mpc}^{1/2}$ can be regarded as upper bounds.  
Our results for $B_{\rm{IG}} = (0.1-1)$ nG and 
${\lambda}_{\rm{coh}} = 1$ Mpc are shown in Table 3. 

\begin{table}[tb]
\begin{center}
\caption{\label{Tab3}Obtained values of $\tau_d^{-1}$ for uniform turbulence IGMFs.}
\begin{tabular}{|c||c|c|}
\hline $s$ & $\tau_d^{-1}$ [$\rm{yr}^{-1}$] for 0.1 nG ${\rm{Mpc}}^{1/2}$ & 
$\tau_d^{-1}$ [$\rm{yr}^{-1}$] for 1.0 nG ${\rm{Mpc}}^{1/2}$\\
\hline
\hline 2.0 & $9.7 \times {10}^{-5}$ & $8.9 \times {10}^{-7}$\\
\hline 2.2 & $9.5 \times {10}^{-5}$ & $8.6 \times {10}^{-7}$\\
\hline 2.4 & $9.3 \times {10}^{-5}$ & $8.4 \times {10}^{-7}$\\
\hline 2.6 & $9.1 \times {10}^{-5}$ & $8.3 \times {10}^{-7}$\\
\hline
\end{tabular}
\end{center}
\end{table}


\section{Implications for Transient UHECR Sources} 
We have estimated lower and upper bounds on $\tau_d$ using 
$\psi \sim 5^{\circ}$, which allows us to estimate the allowed range of 
$\rho_0$ and $\tilde{\mathcal{E}}_{\rm{CR}}^{\rm{iso}}$ 
by using $n_s \sim {10}^{-4} \, \rm{Mpc}^{-3}$, 
As the local rate, we obtain
\begin{equation}
0.1 \, {\rm{Gpc}^{-3} \, \rm{yr}^{-1}} \lesssim \rho_0 \lesssim
(60-3000) \, {\rm{Gpc}^{-3} \, \rm{yr}^{-1}}. 
\end{equation}
Note that stronger upper bounds can be obtained with the GMF with a 
dipole magnetic field. However, since the existence of a dipole field
is very tentative, we hereafter consider the GMF without a
dipole field for conservative discussions.

The required cosmic-ray energy input at ${10}^{19}$ eV, 
$\tilde{\mathcal{E}}_{\rm{HECR}}^{\rm{iso}} 
\equiv \tilde{\mathcal{E}}_{\rm{CR}}^{\rm{iso}}({10}^{19}~{\rm eV})$ is estimated as
\begin{equation}
(0.3-20) \times {10}^{50} \, {\rm{ergs}}  \lesssim 
\tilde{\mathcal{E}}_{\rm{HECR}}^{\rm{iso}} \lesssim {10}^{54} \, {\rm{ergs}}. 
\end{equation}
Note that Eqs. (5) and (6) are valid as long as UHECR sources are 
regarded as transient, i.e., 
$\delta T < \tau_d < \Delta T$, where $\delta T$ is the 
true burst duration during which particle acceleration occurs and 
$\Delta T$ is the time interval between bursts.
$\delta T$ depends on the nature of potential sources. For example, classical
high-luminosity (HL) GRBs have $\delta T \sim {10}^{1-2}$ s, which is
much shorter than $\tau_d$.
For bounds to be meaningful, $\tau_d < \Delta T$ should be satisfied.
Otherwise, more than one UHECR bursts occur in 
$\tau_d$ 
within 
$\psi$, and we would see these bursts as a single but more energetic
burst. When $\psi \sim 5^{\circ}$, the time interval is estimated as 
${\Delta T} \sim (3/\pi) \rho_0^{-1} \psi^{-2} D_{\rm max}^{-3} (E_{\rm th}) 
\simeq 3 \tau_d n_{s,-4}^{-1}$. Since $\Delta T > \tau_d$, we may expect 
that obtained bounds would make a sense, although we should be careful
of the possibility not to see each UHECR burst as a 
distinctive one for larger $n_s$.

The total cosmic-ray energy input $\mathcal{E}_{\rm CR}^{\rm
iso}$ is generally larger than $\tilde{\mathcal E}_{\rm HECR}^{\rm
iso}$ by $R({10}^{19}~{\rm eV}) \equiv (\int d
E_g^{\prime}~E_g^{\prime}~\frac{dN}{d E_g^{\prime}})
/{(E_g^2 \frac{dN}{d E_g})}_{E_g={10}^{19}~{\rm eV}}$ \cite{MINN08}. 
$R$ depends on the cosmic-ray spectrum at a source, and we expect $R
\sim 20-500$ for $s \sim 2.0-2.2$ expected in the ankle scenario while 
$R \gtrsim 100$ for $s \sim 2.4-2.6$ expected in the proton-dip
scenario, and the latter scenario generally requires the break 
energy below the second knee \cite{BGG06}. In both scenarios, we expect 
that the transient hypothesis requires the relatively large 
cosmic-ray energy input per burst 
$\mathcal{E}_{\rm CR}^{\rm iso} \gtrsim {10}^{50.5}~{\rm
ergs}$, which would be a strong requirement on potential sources.
\begin{table}[tb]
\begin{center}
\caption{Potential Sources and Their Typical Local Rates} 
\begin{tabular}{|c||c|c|}
\hline Source & Typical Rate $\rho_0$ [$\rm{Gpc}^{-3} \, \rm{yr}^{-1}$] & Reference\\
\hline
\hline HL GRB & $\sim$ 0.1 & e.g., GP07\\
\hline LL GRB & $\sim$ 400 & e.g., L+07\\
\hline Hypernovae  & $\sim$ 2000 & e.g., GD07\\
\hline Magnetar & $\sim$ 12000 & e.g., G+05\\
\hline Giant Magnetar Flare & $\sim$ 10000 & e.g., O07\\
\hline Giant AGN Flare & $\sim$ 1000 & FG08\\
\hline SNe Ibc & $\sim$ 20000 & e.g., GD07\\
\hline Core Collapse SNe & 120000 & e.g., M+98\\
\hline
\end{tabular}
\end{center}
\end{table}

So far, several potential sources are proposed as transient
accelerators, and HL GRB is one of them. The isotropic radiation energy is 
${\mathcal{E}}_{\gamma}^{\rm{iso}} \sim {10}^{53}$ ergs, and HL GRBs
are the most energetic transient phenomena in the universe.
The local rate is uncertain, but recently suggested rates in the Swift
era, $\rho_0 \sim (0.05-0.27)~{\rm Gpc}^{-3}~{\rm yr}^{-1}$ are 
smaller than previous ones \cite{LD07,GP07}.  
If $\rho_0 \lesssim 0.1~{\rm Gpc}^{-3}~{\rm yr}^{-1}$ is real, 
HL GRBs would be difficult as UHECR sources, since they require 
rather strong IGMFs with $B_{\rm IG} \gtrsim$ nG and large
isotropic energy input of $\mathcal{E}_{\rm CR}^{\rm{iso}} 
\gtrsim 2 \times {10}^{55} (R/20)~{\rm ergs}$. 
Low-luminosity (LL) gamma-ray bursts may overcome the problem that 
the local rate of HL GRBs seems too small \cite{MINN06,MINN08}, 
because their local rate is likely to be much higher, $\rho_0 \sim
{10}^{2-3}~{\rm Gpc}^{-3}~{\rm yr}^{-1}$ \cite{Liaea07,GD07}.
Some GRBs are associated with energetic supernovae (SNe) called
hypernovae, which may also be high-energy cosmic-ray accelerators. Although
they are sufficient as the energy budget, it seems difficult to 
accelerate protons up to $\gtrsim {10}^{19}~{\rm eV}$ \cite{WRMD07}. 

About $10$ \% of core collapse SNe may form magnetars \citep[e.g.,][]{
Gaeea05}, which may be UHECR sources \cite{Aro03}. However, our results 
would suggest that all the magnetars (and SNe) do not produce UHECRs 
uniformly and only a fraction of magnetars is the main origin, 
which is also consistent with the theoretical expectation \cite{Aro03}. 
For example, only newly born magnetars associated with SNe Ibc could be
major UHECR accelerators, which leads to $\rho_0 \sim 3000~{\rm
Gpc}^{-3}~{\rm yr}^{-1}$ and 
$\mathcal{E}_{\rm{CR}}^{\rm{iso}} \sim 3 \times {10}^{50} (R/10)~{\rm ergs}$.
Giant magnetar flares could not explain UHECRs, since their 
radiation energy, $\mathcal{E}_{\gamma}^{\rm iso} \sim {10}^{46}~{\rm ergs}$
is much smaller than $\mathcal{E}_{\rm{CR}}^{\rm iso}$. 

AGNs are the most discussed UHECR accelerators, and the possibility as
transient sources was recently suggested \cite{FG08}. Although such
giant AGN flares may be UHECR sources, the suggestion is speculative. 
If we adopt $\rho_0 \sim {10}^{2-3}~{\rm Gpc}^{-3}~{\rm yr}^{-1}$
although the rate is also uncertain, the required
energy input is $\mathcal{E}_{\rm CR}^{\rm iso} \sim 2 \times {10}^{51-52}
(R/20)~{\rm ergs}$. The corresponding luminosity is 
$L_{\rm{CR}}^{\rm iso} \sim 2 \times {10}^{46-47}~(R/20)~
({10}^{5}~{\rm s}/{\delta T})$ ergs s$^{-1}$. 

The typical source rates are summarized in Table 4, but one
should keep in mind that they are very uncertain at present.

\section{Summary and Discussions} 
We have constrained $\tau_d$ through our numetical calculations of 
proton propagation, taking into account not only the IGMF but also GMF. 
The recent PAO results, the positional correlation with $\psi \sim$ a
few degrees and small-scale anisotropy, have brought us implications for
the sources, as well as indicated $B_{\rm{IG}} {\lambda}_{\rm
coh}^{1/2} \lesssim {\rm nG}~{\rm Mpc}^{1/2}$. 
The results 
would be important in the sense that they are implications of UHECR
observations and also useful for the detectability of secondary emission. 
They suggest that HL GRBs may be marginally disfavored as UHECR
sources, if the recently suggested local rate is real. 
LL GRBs, newly born magnetars associated with LL GRBs or SNe Ibc, and 
giant AGN flares seems possible, 
unless there exists a strong dipole magnetic field in our Galaxy.     

Although these implications may be interesting, because of current poor
stastistics, we should be careful to reach definite conclusions 
about the sources. The suggested posional correlation is confirmed
just at 2-3 sigma levels. Also, the estimate of $n_s$ has 
large errors at present 
due to not only poor statistics but also the lack of our knowledge 
of the precise positions of UHECR sources, 
and it is not so easy to exclude the
possibility that UHECR sources contains many dim accelerators, i.e., 
large $n_s$. 
But, statistics will be better in the near future, and 
the order of $n_s$ will be accurately 
determined by 5 yrs observations by the PAO \cite{TS07}. 
However, statistical analyses become more complicated 
when the sources are transient and/or their luminosity function 
is taken into account (and note that we have assumed 
that the sources are uniform). 
More detailed and careful studies will be presented in our forthcoming paper.

We have discussed implications for the transient UHECR sources, 
but it is also important to know whether the sources are transient or not. 
The signature of transient sources may be found from the observed UHECR
spectrum, e.g., from the average number of multiplets \cite{HMR04}.
For identification of sources, the multimessenger astronomy will be 
particularly important. High-energy neutrinos are useful as a probe of 
cosmic-ray acceleration, and associated gamma rays are often expected for
transient sources such as GRBs 
\citep[see, e.g.,][and references there in]{M07,MINN08}.
In addition, electrons are also accelerated in the shock acceleration
theory, which allows us to expect photon counterparts. In fact, we may
also expect the high electron luminosity $L_e \gtrsim (\epsilon_e/0.1
\epsilon_p){10}^{45} ({10}^{5}~{\rm s}/ \delta T)$ ergs s$^{-1}$, as well as
the high magnetic luminosity $L_B \gtrsim {10}^{45}~{\rm ergs~s^{-1}}$
which will be required for UHECR acceleration \cite{Wax95a}.
For example, $\sim$ 30 events from giant AGN flares may be detected by
Fermi \cite{FG08}.
Note that, once we know that the UHECR sources are transient, we can
obtain precious information on the effective IGMF \cite{MW96}. 
If LL GRBs are the UHECR sources, for example, the effective IGMF 
can be estimated as $\sim 0.03~{\rm nG}~{\rm Mpc}^{1/2}$. 
Also, UHECRs would also be useful as a probe of the GMF \cite{TS08a}, and 
several GMF models may be tested as the number of detected events
increases. This can help to reduce uncertainties in estimate of bounds on 
$\tau_d$ and resulting $\rho_0$.

We have assumed an uniform IGMF, but it would not be realistic.
If the sources are inside clusters or filaments, 
UHECRs are affected by the magnetic field in the structured region. 
When an uniform IGMF is weak enough, such the structured magnetic field
is more important. In such cases, HL GRBs are more disfavored as the UHECR
sources because they require the rather strong IGMF, as shown
in this work. The structured magnetic field may also play a role as an
unavoidable field. Some AGNs are inside clusters, so that UHECRs from
them should be delayed due to that field. The local rate will be 
more constrained as $\rho_0 \lesssim {10}^{1-2}~{\rm
Gpc}^{-3}~{\rm yr}^{-1}$, and the necessary cosmic-ray
energy input will also be increased. 
 


\acknowledgments
We thank K. Sato, S. Nagataki, T. Tanaka, N. Seto,
T. Nakamura, K. Ioka and S. Inoue for useful 
comments. We are supported by the JSPS fellowship.
Support also comes from 
the 
GCOE Program
"The Next Generation of Physics, Spun from Universality and  
Emergence" from MEXT.


\clearpage





\end{document}